\documentclass[conference]{IEEEtran}
\IEEEoverridecommandlockouts
\usepackage{cite}
\usepackage{amsmath,amssymb,amsfonts}
\usepackage{graphicx}
\usepackage{textcomp}
\usepackage{xcolor}
\usepackage{graphicx}
\usepackage{subcaption}
\usepackage{siunitx}  
\usepackage{booktabs}
\usepackage{algorithm}
\usepackage{algpseudocode}
\usepackage{tikz}
\def\BibTeX{{\rm B\kern-.05em{\sc i\kern-.025em b}\kern-.08em
    T\kern-.1667em\lower.7ex\hbox{E}\kern-.125emX}}
\begin{document}

\title{Underwater Acoustic Signal Recognition 
\\Based on Salient Features}

		\author{
		\IEEEauthorblockN{Yang Zhang}
		\IEEEauthorblockA{$^1$\textit{College of Information Science and Technology},\\
			Beijing University of Chemical Technology,\\
			Beijing, China.\\
			$^2$\textit{Zhongfa Aviation Institute},\\
			Beihang University,\\
			Hangzhou, China.\\
			yang\_zh@mail.buct.edu.cn}\\

		\IEEEauthorblockN{Minghao Chen}
		\IEEEauthorblockA{$^1$\textit{College of Information Science and Technology},\\
			Beijing University of Chemical Technology,\\
			Beijing, China.\\
			2023200799@mail.buct.edu.cn}\\
		
		\IEEEauthorblockN{Yuan Xu*}
		\IEEEauthorblockA{$^1$\textit{College of Information Science and Technology},\\
			Beijing University of Chemical Technology,\\
			Beijing, China.\\
			xuyuan@mail.buct.edu.cn
		}

		\and
		
			\IEEEauthorblockN{Zhizhen Wang}
		\IEEEauthorblockA{\textit{College of Information Science and Technology},\\
			Beijing University of Chemical Technology,\\
			Beijing, China.\\
			2023200830@mail.buct.edu.cn}\\

		\IEEEauthorblockN{Yanlin He}
		\IEEEauthorblockA{$^1$\textit{College of Information Science and Technology},\\
			Beijing University of Chemical Technology,\\
			Beijing, China.\\
			heyl@mail.buct.edu.cn
		}\\
		
		\IEEEauthorblockN{Qunxiong Zhu}
		\IEEEauthorblockA{$^1$\textit{College of Information Science and Technology},\\
			Beijing University of Chemical Technology,\\
			Beijing, China.\\
			zhuqx@mail.buct.edu.cn
		}
		
	}

\maketitle

\begin{abstract}
With the rapid advancement of technology, the recognition of underwater acoustic signals in complex environments has become increasingly crucial. Currently, mainstream underwater acoustic signal recognition relies primarily on time-frequency analysis to extract spectral features, finding widespread applications in the field. However, existing recognition methods heavily depend on expert systems, facing limitations such as restricted knowledge bases and challenges in handling complex relationships. These limitations stem from the complexity and maintenance difficulties associated with rules or inference engines. Recognizing the potential advantages of deep learning in handling intricate relationships, this paper proposes a method utilizing neural networks for underwater acoustic signal recognition. The proposed approach involves continual learning of features extracted from spectra for the classification of underwater acoustic signals. Deep learning models can automatically learn abstract features from data and continually adjust weights during training to enhance classification performance. 
\end{abstract}


\section{Introduction}

With the rapid advancement of technology, the importance of recognizing underwater acoustic signals in complex environments has become increasingly significant. Current mainstream methods for underwater acoustic signal recognition heavily rely on time-frequency analysis to extract spectrogram features, which have widespread applications in the field of underwater signal recognition. Presently, mainstream underwater signal recognition primarily relies on expert systems, leading to limitations such as knowledge base constraints and difficulties in handling complex relationships. These limitations stem from the complexity and maintainability challenges associated with rules or inference engines. Recognizing the potential advantages of deep learning methods in handling complex relationships, this paper proposes the use of neural network methods for underwater acoustic signal recognition.

The proposed approach involves continuously learning and classifying underwater acoustic signals based on features extracted from spectrograms. Deep learning models have the capability to automatically learn abstract features from data, continuously adjusting weights during training to enhance classification performance. The process is illustrated in the flowchart shown in Figure 1.

The contributions of this paper lie in:

1. Introducing an automated method for extracting Demon spectrogram features.
2. Proposing a method for classifying ship types using deep learning.

These contributions aim to advance the field of underwater signal recognition by automating feature extraction and leveraging the capabilities of deep learning for more effective classification of underwater acoustic signals.

\begin{figure}
\centerline{\includegraphics[scale=0.35]{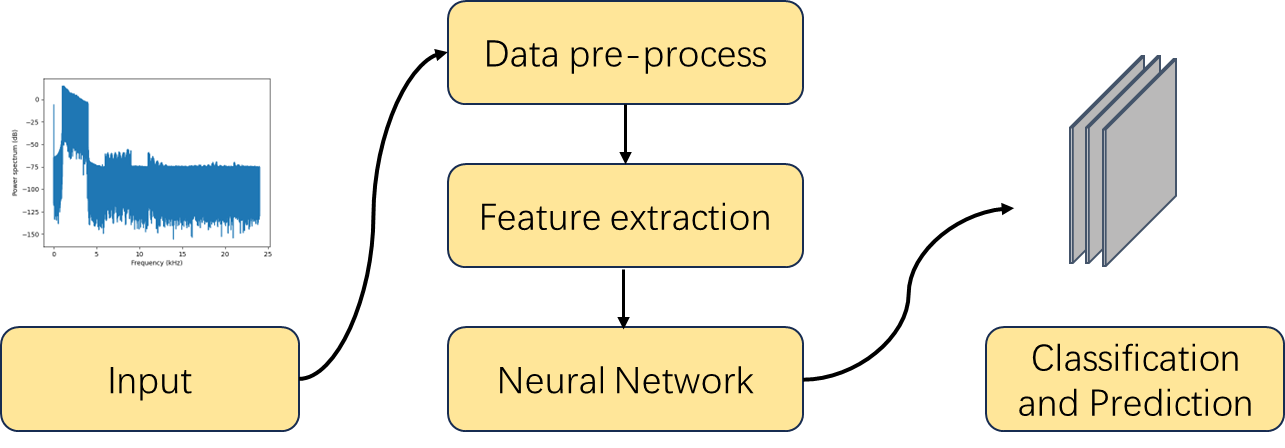}}

\caption{Flowchart of the algorithm}
\label{fig: flowchart}
\end{figure}

\section{Related Work}

In recent years, significant progress has been made in various aspects of underwater acoustic signal processing, giving rise to a series of impactful studies. Firstly, the groundbreaking research by Tao\cite{Tao2007}delved into the spectral analysis and reconstruction of periodically nonuniformly sampled signals. Introducing the fractional Fourier domain, they proposed an efficient method for effectively restoring spectral information, laying a solid foundation for signal analysis in the field of underwater acoustic signal processing.

Subsequently, Hinton \cite{Hinton2012}introduced the application of deep neural networks in speech recognition, showcasing its robust performance in acoustic modeling. This innovation introduced machine learning to the realm of underwater acoustic signal processing, establishing a framework for subsequent research. The introduction of deep neural networks provided a new paradigm for underwater acoustic signal processing, sparking interest in automated processing.

Wei \cite{Wei2018} proposed a deep neural network-based method for underwater acoustic signal classification, opening new avenues for automated signal categorization. This method not only improved classification accuracy but also injected more elements of machine learning into the field, driving its development towards intelligence.

In their 2023 study, Xu\cite{Xu2023}successfully enhanced the accuracy of underwater acoustic signal classification through self-supervised learning and the Mixup technique. The introduction of this method emphasizes the potential role of data augmentation and self-supervised learning in underwater acoustic signal processing, providing a novel approach to improving classification performance.

The research by Miao\cite{Miao2021}employed sparse time-frequency representation and deep learning methods, offering an innovative approach to underwater acoustic signal classification. This combination of traditional signal processing techniques and deep learning enriched the technical means of underwater acoustic signal processing, providing more choices for handling complex signals.

Zhang's work\cite{Zhang2016} in 2016 extensively explored the application of deep learning in remote sensing data processing, expanding the scope of deep learning beyond underwater acoustic signal processing. This indicates the broad potential of deep learning technology, not limited to specific fields, offering insights for interdisciplinary applications in underwater acoustic signal processing.

In Domingos's review in 2022\cite{Domingos2022}, various methods using deep learning for underwater data classification were comprehensively summarized, with a particular focus on shoreline surveillance. This review provides researchers with a comprehensive understanding of the overall development direction of underwater data processing, offering important guidance for future research.

Lastly, Ma\cite{Ma2022}addressed the spectrum analysis for multiband signals under the context of nonuniform sub-Nyquist sampling. This study overcame challenges in spectrum analysis under specific sampling conditions, providing robust support for handling complex signals in real-world scenarios.

In summary, these studies collectively reveal a diverse range of techniques in underwater acoustic signal processing, from traditional signal analysis methods to the application of deep learning. They offer a broad reference for future research. These advancements not only drive the theoretical development of underwater acoustic signal processing but also provide rich tools and methods for practical applications. As technology continues to evolve, we can expect deeper breakthroughs in the field of underwater acoustic signal processing.

\section{Method}
\subsection{Demon Spectrum Line Feature Extraction}\label{AA}

\begin{figure}[b]
\centerline{\includegraphics[scale=0.2]{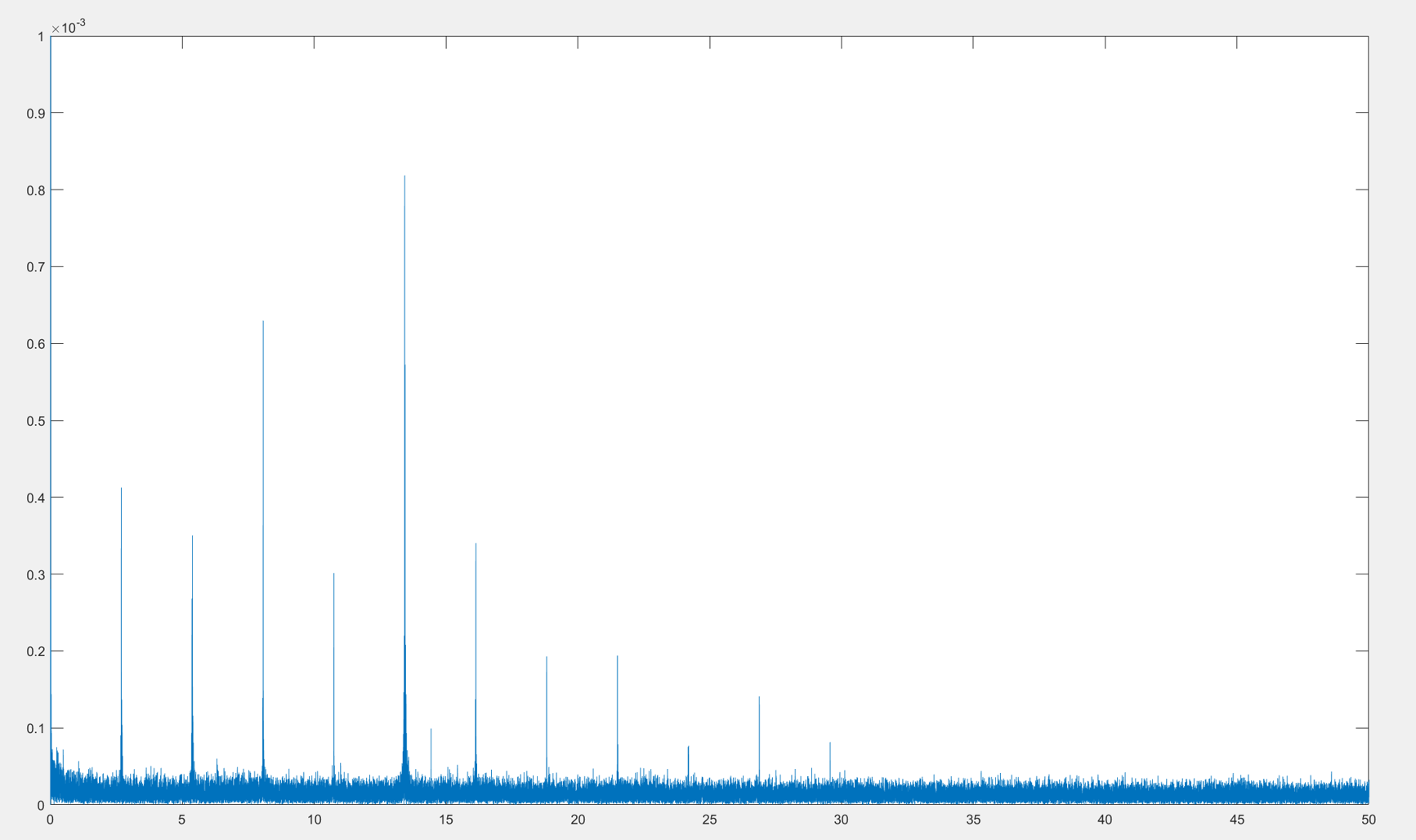}}

\caption{Demon spectrum}
\label{fig:device}
\end{figure}

The conventional Demon spectrum line \cite{Tao2007}\cite{Ma2022}is a method of visualizing signal spectra by processing spectrograms to generate a color intensity distribution plot. It is primarily used for intuitively presenting the characteristics of signals in the time-frequency domain. To achieve automated recognition of underwater acoustic signals, this study has designed a system for the automated extraction of Demon features, converting Demon spectrum line plots into feature vectors suitable for neural network input.The visual results of the Demon spectrum are shown in Fig\ref{fig:device}.

This feature vector comprises five dimensions, including blade frequency, shaft frequency, average wave strength, maximum shaft frequency, and maximum blade frequency. Each dimension reflects key features of underwater acoustic signals in the time-frequency domain. Shaft frequency and blade frequency are the primary criteria for determining vessel types using Demon spectrum lines since different types of vessels often have varying numbers of blades, e.g., fishing vessels tend to have fewer blades compared to other surface vessels. Additionally, due to lower power, fishing vessels often exhibit faster rotation speeds.

The average wave strength reflects the overall intensity level of Demon spectrum lines across the entire frequency spectrum. By calculating the average strength of the waves, we can obtain information about the energy distribution of the overall spectrum, providing the neural model with information about the overall signal intensity. Since the frequency strength of Demon spectrum lines varies significantly for different types of vessels, the average frequency strength holds valuable reference value.

Furthermore, we extract the maximum shaft frequency and maximum blade frequency as additional inputs to the feature vector. The maximum shaft frequency represents the shaft frequency value with the highest spectral intensity in Demon spectrum lines, indicating the position of the dominant frequency in the audio signal. In signal processing, the dominant frequency is usually the most significant part of the signal, corresponding to its core component. From a signal-to-noise ratio perspective, the intensity of the dominant frequency is typically more pronounced than noise, contributing to accurate localization of the dominant frequency in complex signals and improving the separation of signal and noise. The maximum blade frequency reflects the blade frequency value with the highest spectral intensity in Demon spectrum lines, representing the position of the most significant frequency component in the signal. In high-noise environments, noise often spreads across multiple positions in the spectrum, potentially masking important components of the signal. By extracting the maximum blade frequency, we can more sensitively capture the strongest frequency distribution in the signal, reducing the impact of noise and improving signal recognition and signal-to-noise ratio.

\subsection{Neural Network Model}\label{AA}

\begin{figure}[b]
\centerline{\includegraphics[scale=0.4]{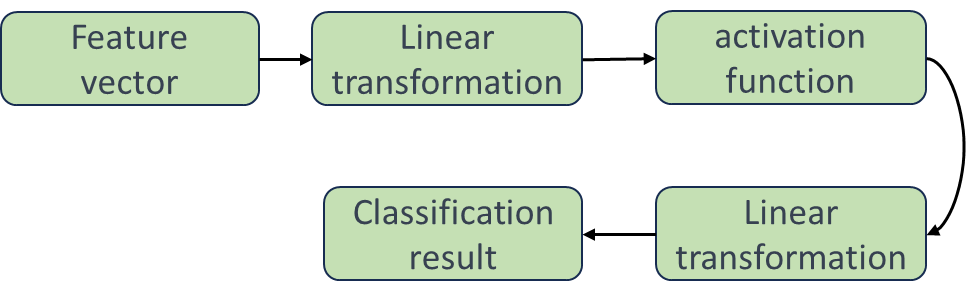}}
\caption{structure of the module}
\label{fig:structure}
\end{figure}

%

Due to the relatively small dataset, there is a potential issue of overfitting in the model, making it less applicable to larger-scale datasets. Additionally, our observation suggests that neural network models tend to excel in predicting the correct category compared to predicting specific types.

Taking these considerations into account, we leverage these observations as prior knowledge to guide the design of a neural network. To address these challenges, we propose a cascaded neural network model. This model operates on the principle that it first predicts the broad category of ships and subsequently refines its prediction to the specific type of ship. It's noteworthy that the model structures for four-class and ten-class classifications are analogous, as illustrated in the diagram.

Recognizing the value of using prior knowledge from large-category classification to guide small-category classification, we opt for a cascading approach involving two simple neural network models. This aims to enhance the overall predictive accuracy of the entire system. The algorithmic process of this cascade is depicted in the following Fig\ref{fig:structure}.

With this design, we anticipate mitigating overfitting on a small dataset and improving the robustness of the entire system when dealing with larger datasets. This cascaded model not only possesses enhanced category prediction capabilities but is also likely to perform well in fine-grained classification of ship types. This section provides insights into the rationale behind our model design and its potential advantages, laying the groundwork for the subsequent sections of the report.

\begin{algorithm}
\caption{Neural network based on Cascade method}
\begin{algorithmic}[1]
    \State \textbf{input:} Feature Vector
    \State \textbf{Output:} Vessel Category \& Vessel Model

    \State category = Classification Network 1 \Comment{5cls:0,1,2,3,4}
    \If{category == 0}
        \State vessel category = 0
    \ElsIf{category == 1}
        \State shipType =classification network 2 \Comment{10cls:0-10}
    \ElsIf{category == 2}
        \State vessel category = 2
    \ElsIf{category == 3}
        \State vessel category = 3
    \ElsIf{category == 4}
        \State vessel category = 4
    \EndIf
\end{algorithmic}
\end{algorithm}

\section{Experiment}

\subsection{Dataset}
This study utilizes a private dataset comprising 300 samples from underwater audio monitoring. Each data entry consists of a 20-minute audio recording that has undergone detailed manual annotation, covering rich information such as the type and model of the vessels. Given the sensitivity of audio data, we emphasize the protection of privacy and security for the dataset. Various measures, including anonymization and access control, have been implemented to ensure the legality and security of data usage.

During model training, the entire dataset is partitioned into training and validation sets in a ratio of 8:2, with the additional requirement that each category is represented in the testing set. To ensure the trained model possesses robust predictive capabilities, the model with the best performance on the testing set is saved as the final outcome.

\subsection{the comparison between diffierence module}
%
%

\begin{figure}
  \centering

  \begin{subfigure}{0.4\textwidth}
    \includegraphics[width=\linewidth]{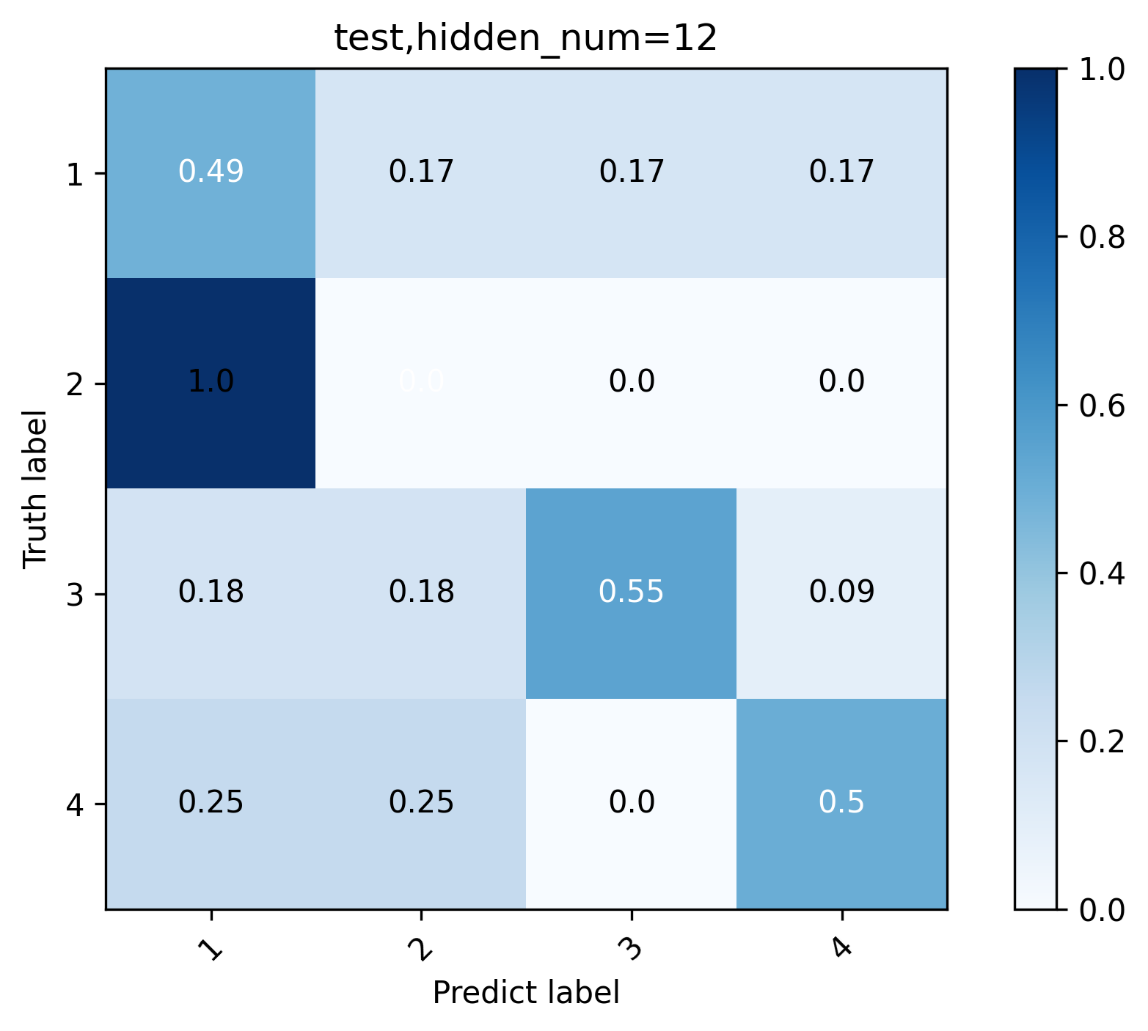}
	\caption{fig:cls4 with 12 hidden number}
  \end{subfigure}
  \begin{subfigure}{0.4\textwidth}
    \includegraphics[width=\linewidth]{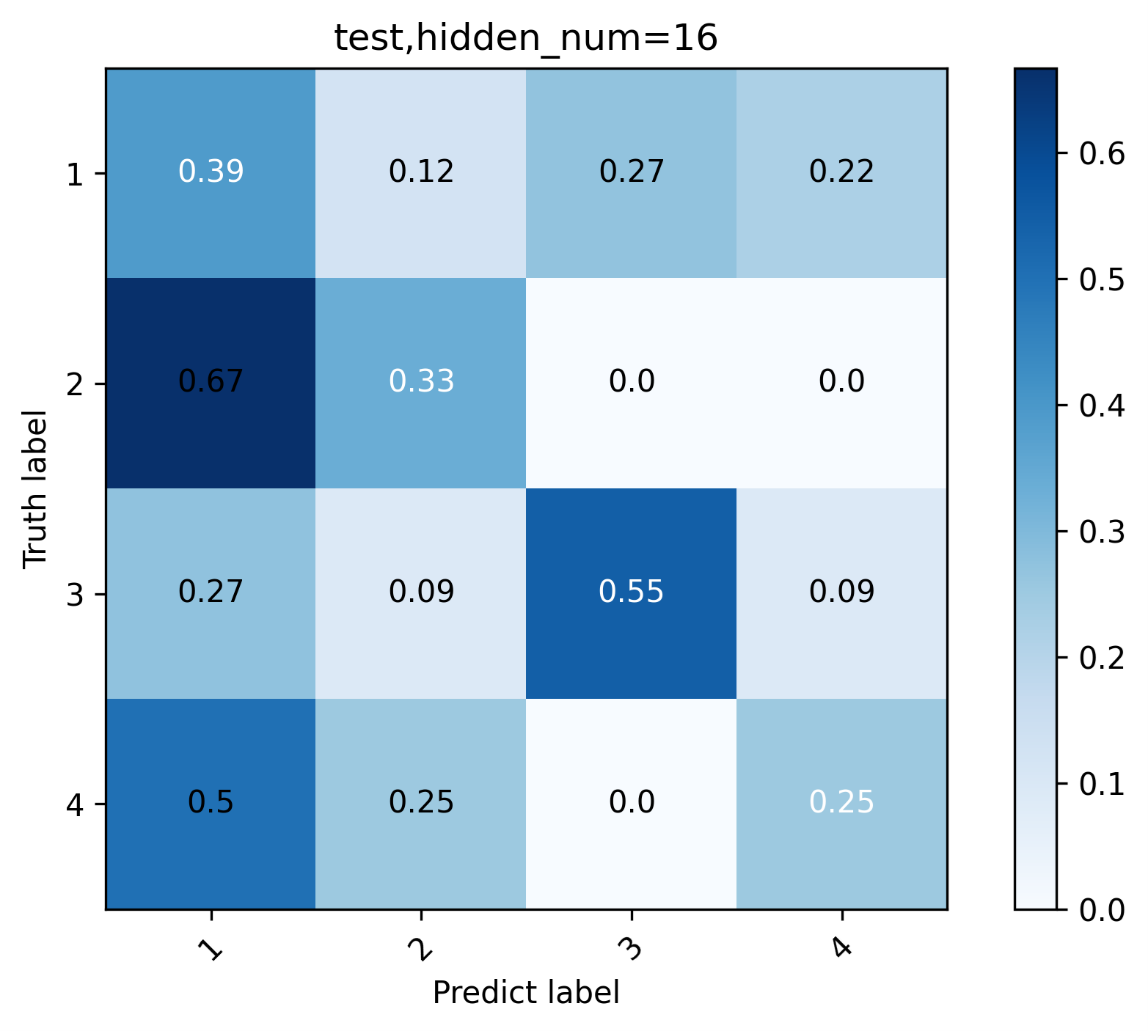}
	\caption{fig:cls4 with 16 hidden number}
  \end{subfigure}
  \begin{subfigure}{0.4\textwidth}
    \includegraphics[width=\linewidth]{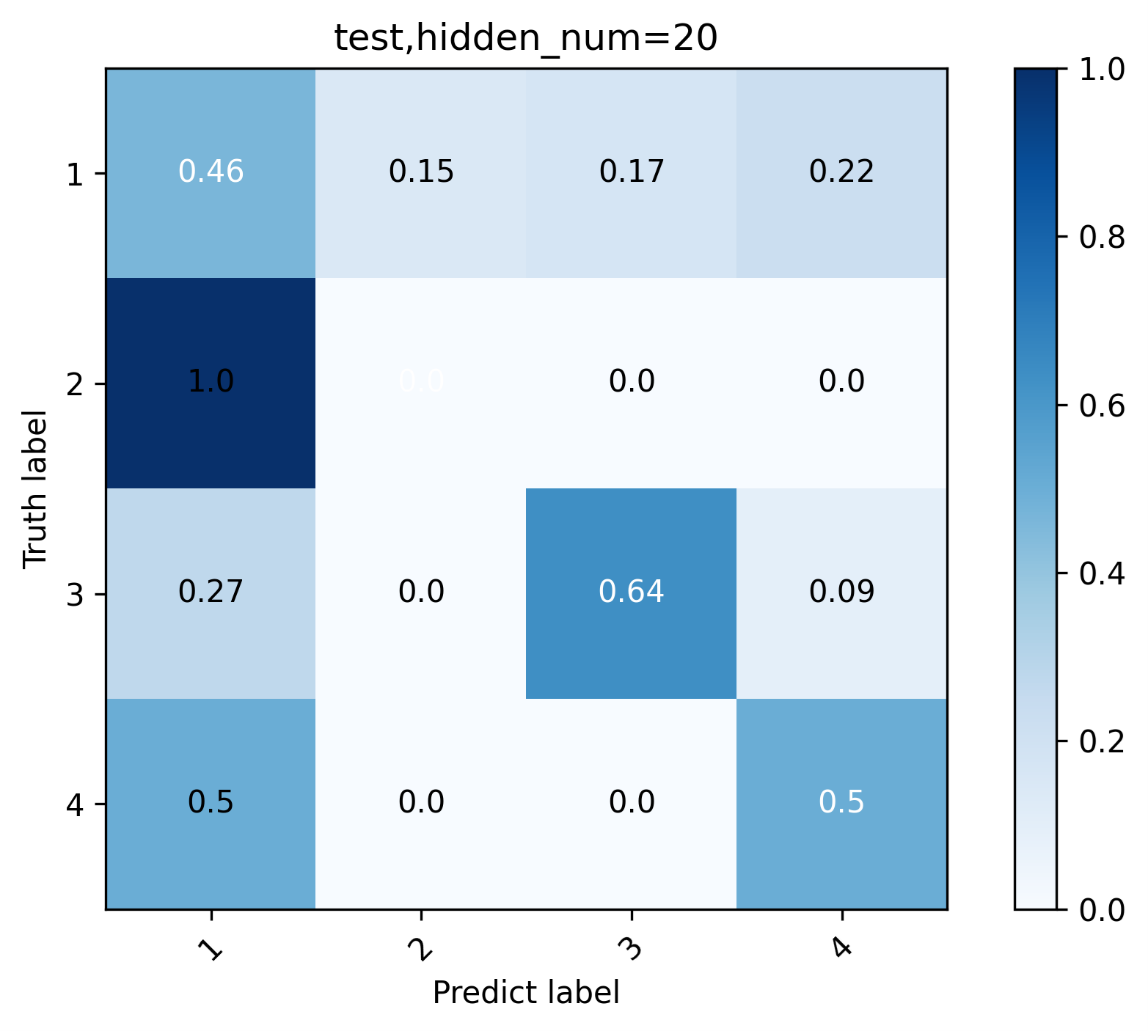}
	\caption{fig:cls4 with 20 hidden number}
  \end{subfigure}

\caption{The results of the 4-class confusion matrix }
\label{res4 }
\end{figure}

\begin{table}[bh]
\begin{center}
\caption{Comparison between diffierent model for 4cls}
\renewcommand\arraystretch{1.2}
\begin{tabular}{c c c c c}
  \toprule[1.2pt]
  The number of hidden layers &  first-cls  &  second-cls  &  third-cls &  fourth-cls  \\
 \hline
  12 & 0.49 & 0 & 0.55 & 0.5 \\
\cline{2-4}  
\hline
  16 &0.39 & 0.33 & 0.55 &0.25 \\
\cline{2-4}  
\hline
  20 & 0.46 & 0 & 0.64 &0.5 \\
  \bottomrule[1.2pt]
\end{tabular}

\label{tab1}
\end{center}
\end{table}

\begin{table}[bh]
\begin{center}
\caption{Comparison between diffierent model for 10cls }
\renewcommand\arraystretch{1.2}
\begin{tabular}{c c c c c}
  \toprule[1.2pt]
  The number of hidden layers &  12 layers  &  16 layers  &  20 layers &  28 layers  \\
\cline{2-4}  
\hline
  cls-1 & 0 & 0 & 0.17&0.33 \\
\cline{2-4}  
\hline
  cls-2 & 0 & 0 & 0 &0\\
\cline{2-4}  
\hline
  cls-3 &0.67 & 0.67 & 0.67&0.33 \\
\cline{2-4}  
\hline
   cls-4 & 0 &0 & 0 &0\\
\cline{2-4}  
\hline
   cls-5 & 0.5 & 0.5 & 0.5&0.5 \\
\cline{2-4}  
\hline
   cls-6 & 0 & 0 & 0& 0 \\
\cline{2-4}  
\hline
   cls-7 & 0.33 & 0.33 &0.33 &0.67\\
\cline{2-4}  
\hline
   cls-8 & 0 & 0 & 0 &0\\
\cline{2-4}  
\hline
   cls-9 & 0.33 & 0.67 & 0.33 &0.33 \\
\cline{2-4}  
\hline
   cls-10 & 0 & 0 & 0  &0\\
  \bottomrule[1.2pt]
\end{tabular}

\label{tab1}
\end{center}
\end{table}


The experiment primarily compared the impact of different neural network structures on model accuracy and predictive capabilities, with the results depicted in the accompanying graph. We investigated the influence of various network structures on model accuracy, focusing on adjustments to the fully connected layer model architecture. The confusion matrix for the experiments is illustrated in the graph. The best results for both the 4-class and 10-class classifications were selected, ensuring comparability by employing the same test set for each experiment.

The experimental findings underscore the effectiveness of the proposed approach in predicting unknown types and models of ships. Notably, the method demonstrated superior performance, as evidenced by its entry into the finals of a prominent algorithmic competition.

These results highlight the robustness and generalizability of the proposed method across diverse ship categories and types, showcasing its potential applicability in real-world scenarios. The utilization of a standardized test set ensures a fair and comprehensive evaluation of the model's performance under different structural configurations. The presented experiment and its outcomes contribute valuable insights to the understanding of the chosen neural network architectures and their implications for ship classification tasks.
%
%
%
%
%

\begin{figure}
  \centering

  \begin{subfigure}{0.35\textwidth}
    \includegraphics[width=\linewidth]{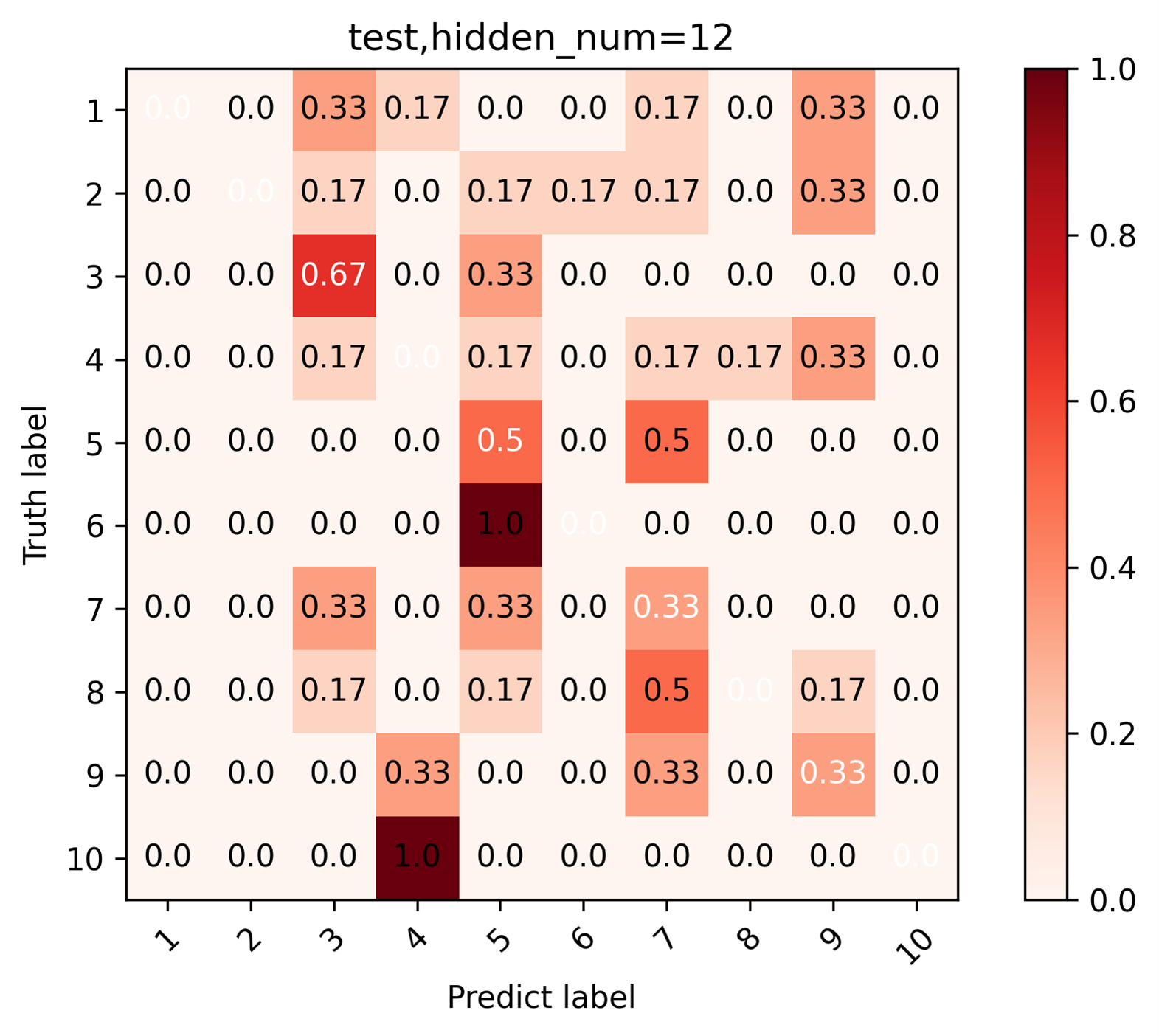}
	\caption{fig:cls10 with 12 hidden number}
  \end{subfigure}
  \begin{subfigure}{0.35\textwidth}
    \includegraphics[width=\linewidth]{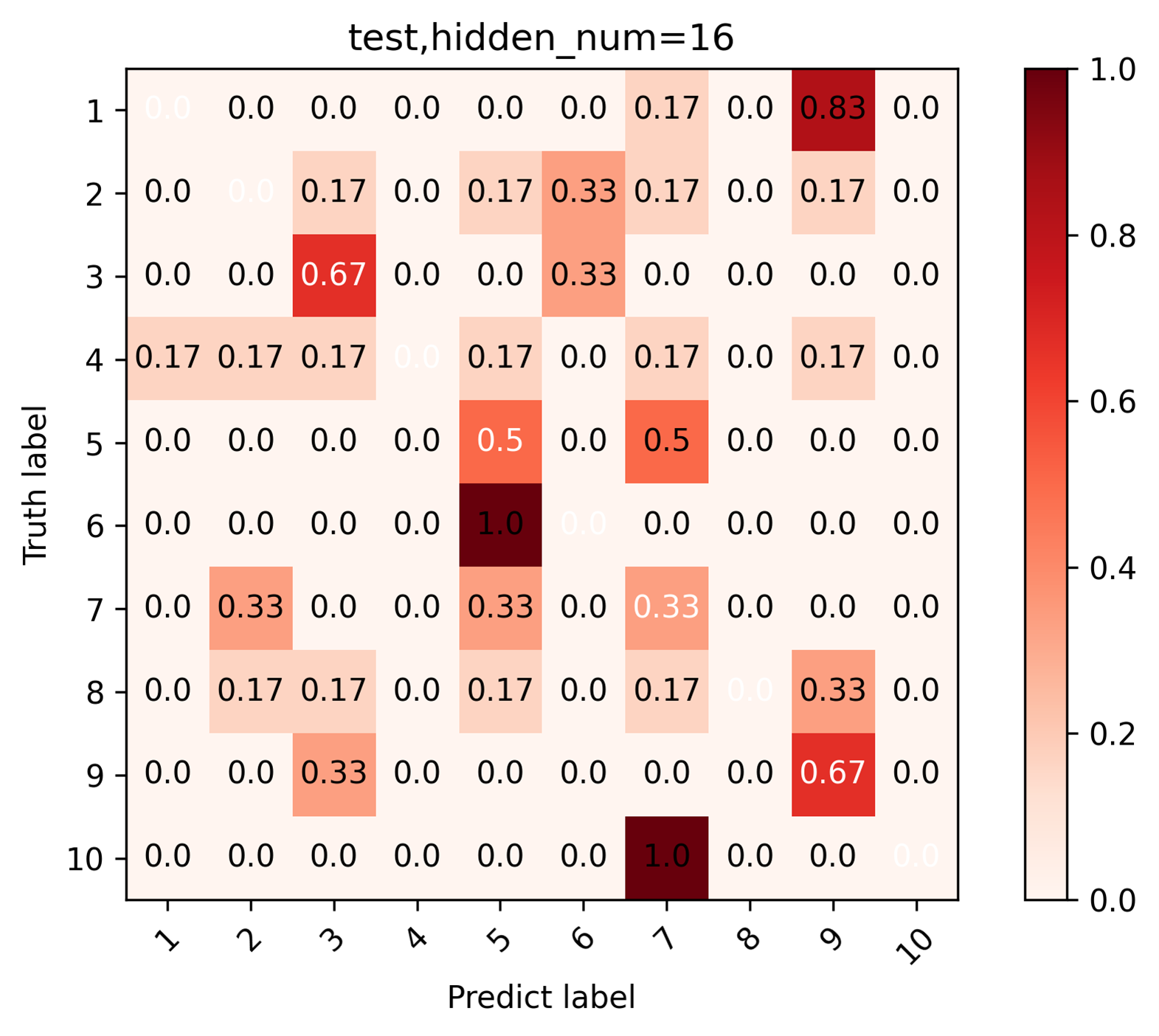}
	\caption{fig:cls10 with 16 hidden number}
  \end{subfigure}
  \begin{subfigure}{0.35\textwidth}
    \includegraphics[width=\linewidth]{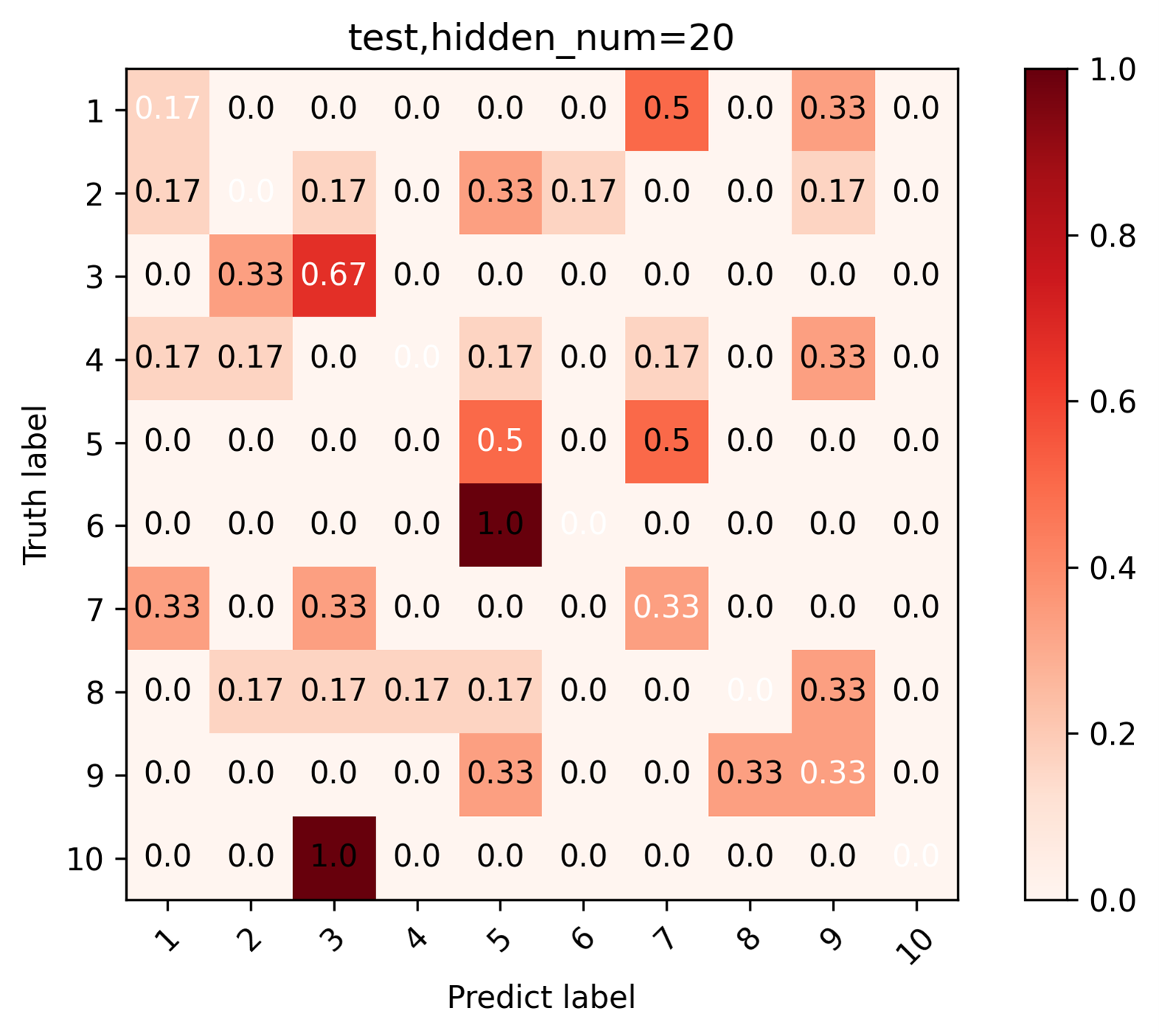}
	\caption{fig:cls10 with 20 hidden number}
  \end{subfigure}
   \begin{subfigure}{0.35\textwidth}
    \includegraphics[width=\linewidth]{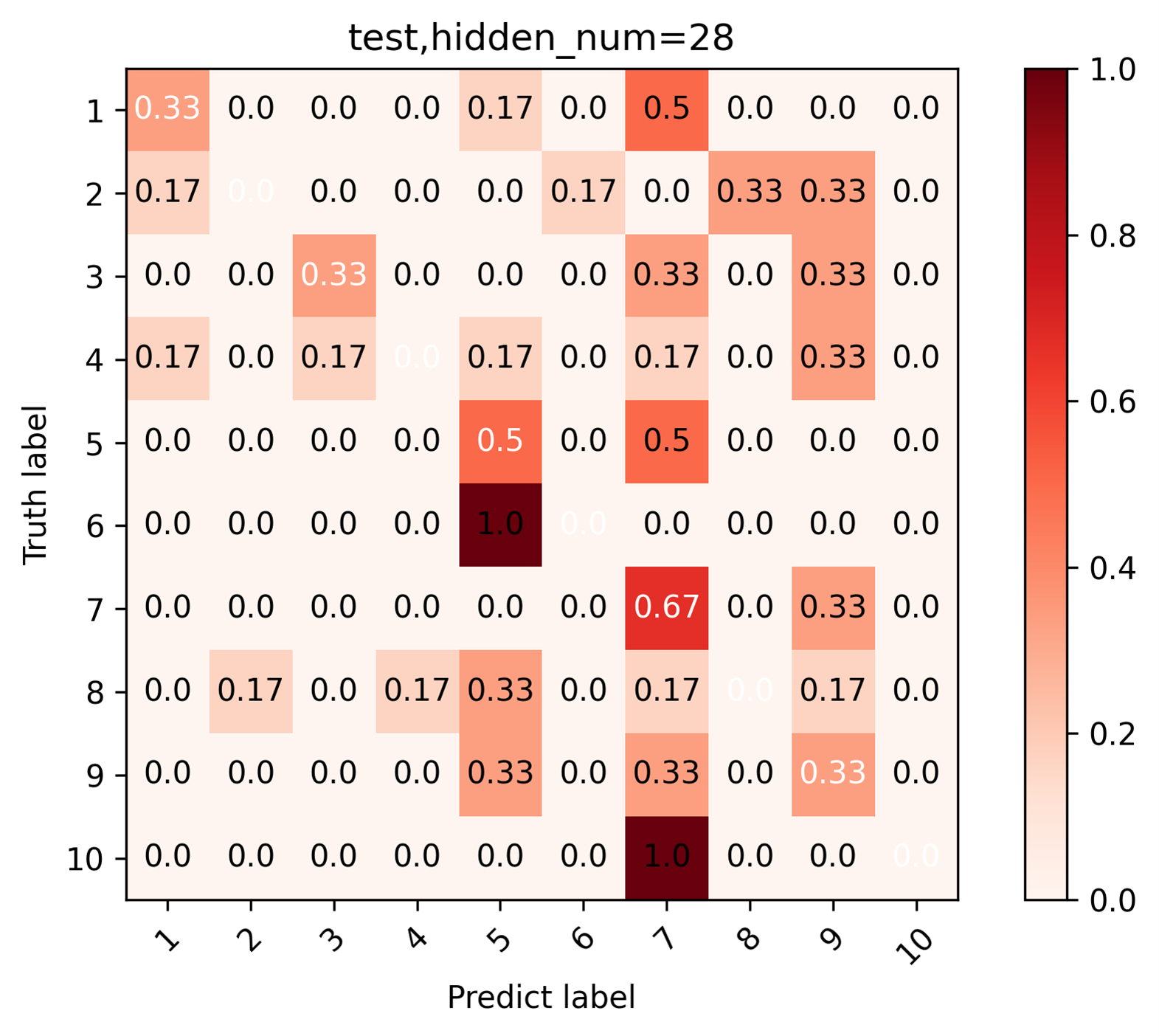}
	\caption{fig:cls10 with 28 hidden number}
  \end{subfigure}

\caption{The results of the 10-class confusion matrix }
\label{res10}
\end{figure}

\section{Conclusion}
%
%
%

This paper explores an effective classification approach in the field of underwater acoustic signal processing based on Demon spectrogram feature extraction and neural networks. Through the Demon spectrogram line extraction method, the comprehensive time-frequency characteristics of underwater acoustic signals are captured. Subsequently, a deep neural network model is introduced, taking the generated Demon spectrogram as input. Through fully connected layers, the model achieves automatic feature learning and classification of underwater acoustic signals. The paper delves into deep learning methods in the field of underwater acoustic signal processing, providing valuable insights for future research in this domain.

The outcomes of this study offer a novel perspective for accurately classifying underwater acoustic signals in complex environments, providing robust support for further developments in related fields. Future considerations may include:

1.Data Augmentation: Segmenting and augmenting audio data to increase the dataset size and improve the model's generalization performance.

2.Multi-level Cascading: Exploring a multi-level cascading approach, initially performing binary classification to enhance accuracy and subsequently conducting finer-grained classification, aiming for higher accuracy at a more granular level.

These future considerations are expected to further optimize the classification methods for underwater acoustic signals, making them more applicable to complex and dynamic real-world scenarios, thereby advancing the field.

{\small
\bibliographystyle{IEEEtran}
\bibliography{TT}
}
\begin{IEEEbiography}[{\includegraphics[width=1in,height=1in,clip,keepaspectratio]{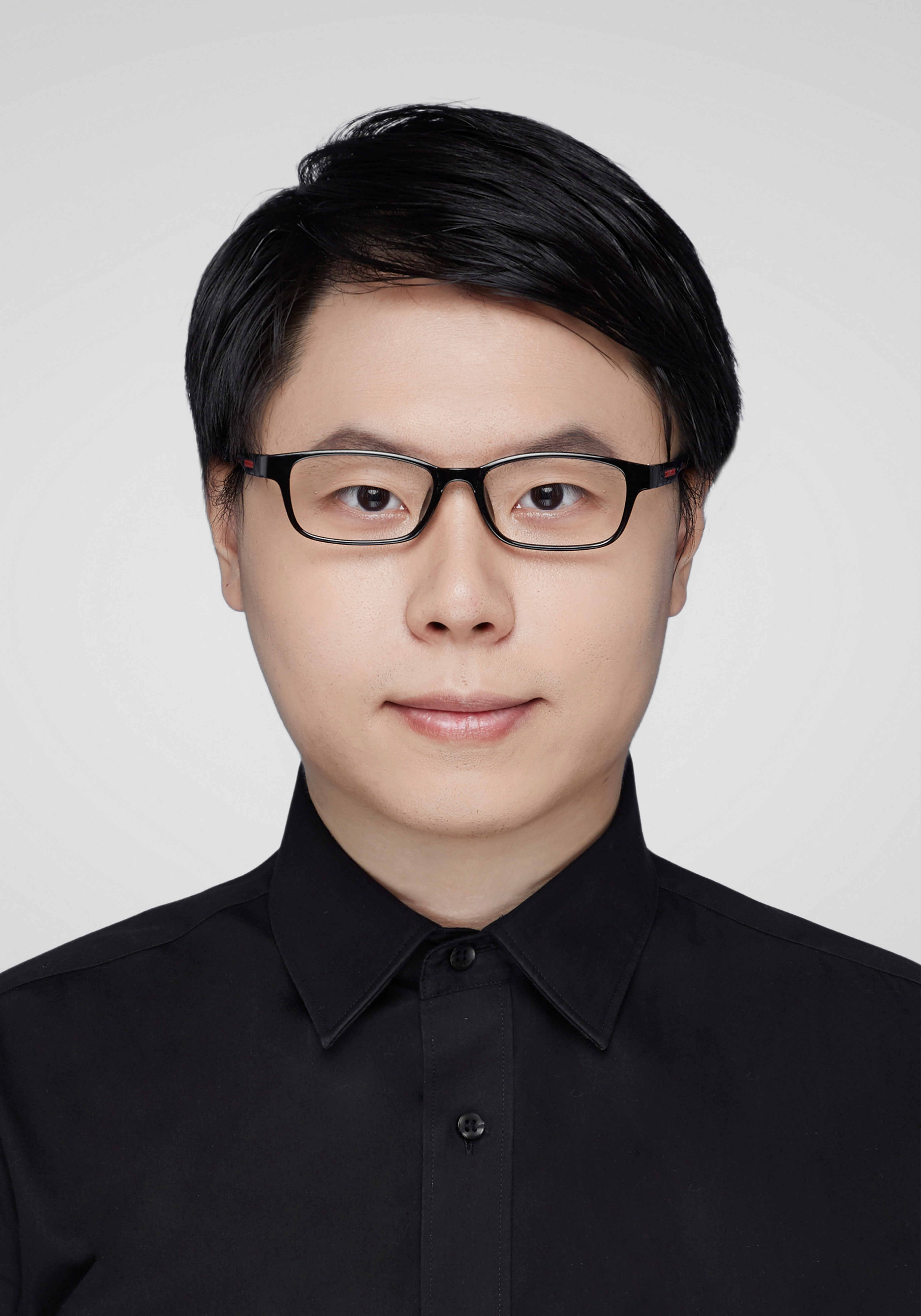}}]{Yang Zhang} received his B.S. and Ph.D. degrees from School of Computer Science and Engineering of Beihang University in 2014 and 2020. He is the corresponding author of this paper. Contact him at yang\_zh@mail.buct.edu.cn. 

Now he is an associate professor in the College of Information Science and Technology, Beijing University of Chemical Technology, China. He is working on computer vision and machine learning.
\end{IEEEbiography}

\begin{IEEEbiography}[{\includegraphics[width=1in,height=1in,clip,keepaspectratio]{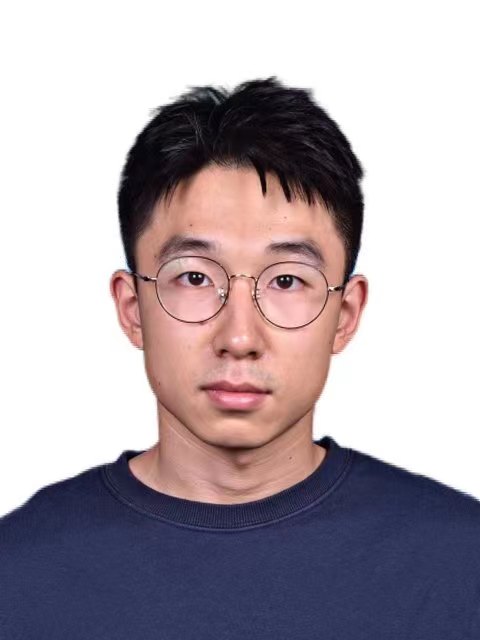}}]{Zhizhen Wang}  received the B.S degree in College of Information Science and Technology from Beijing University of Chemical Technology in 2023. Now he is working toward the master's degree in College of Information Science and Technology from Beijing University of Chemical Technology, Beijing, China. 
\end{IEEEbiography}

\begin{IEEEbiography}[{\includegraphics[width=1in,height=1in,clip,keepaspectratio]{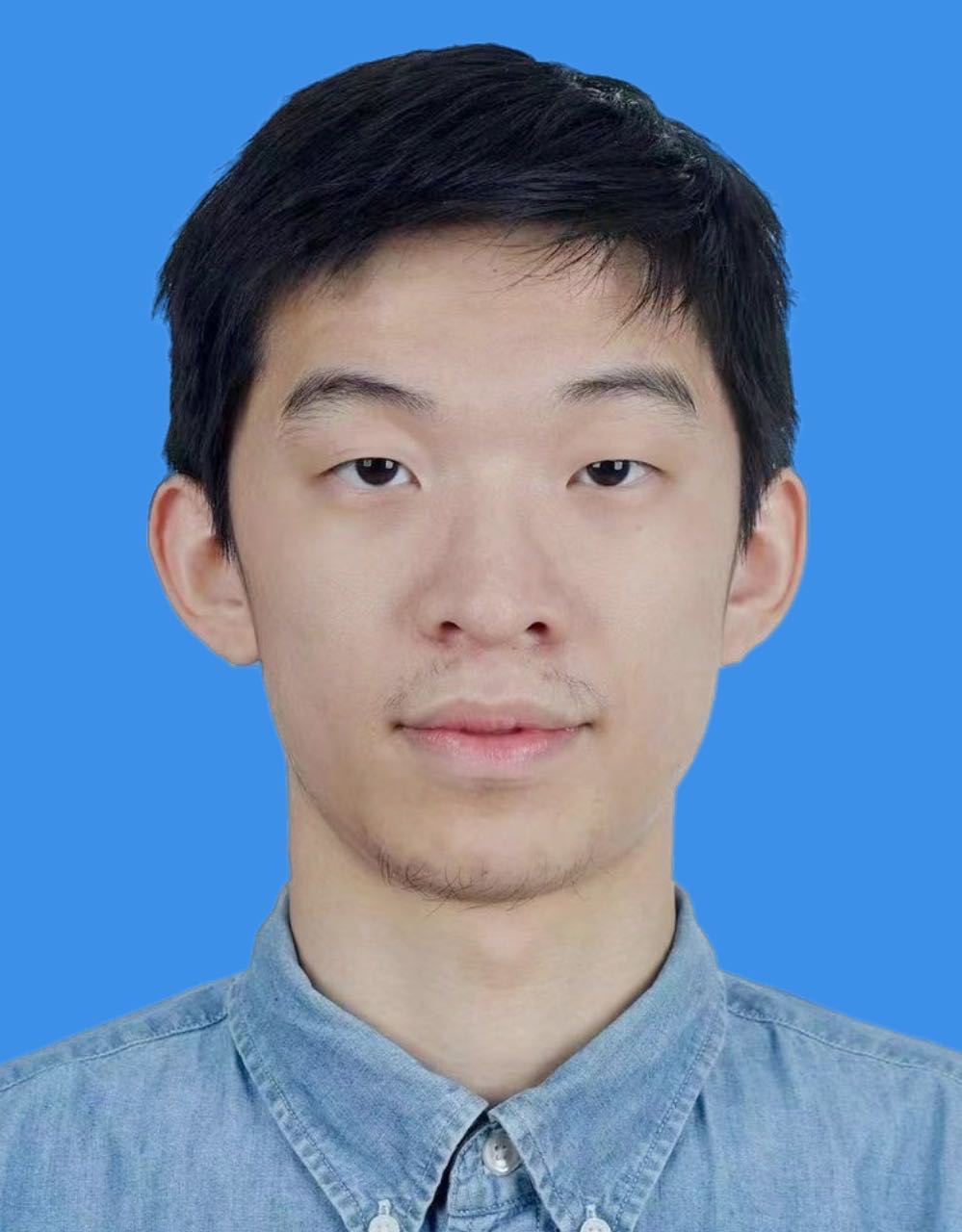}}]{Minghao Chen} received the B.S degree in College of Information Science and Technology from Beijing University of Chemical Technology in 2023. Now he is working toward the master's degree in College of Information Science and Technology from Beijing University of Chemical Technology, Beijing, China. 

Now his research interest is computer vision, and he is particularly interested in object detection.
\end{IEEEbiography}

\begin{IEEEbiography}[{\includegraphics[width=1in,height=1in,clip,keepaspectratio]{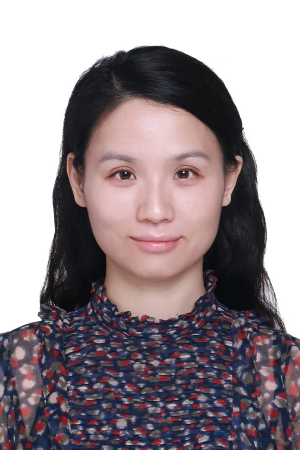}}]{Yuan Xu} received the B.S. and Ph.D. degrees in College of Information Science and Technology from Beijing University of Chemical Technology, in 2005 and 2010, respectively. 

Now she is a professor in College of Information Science and Technology from Beijing University of Chemical Technology, Beijing, China. Her research interests include computational intelligence, data mining, fault diagnosis, machine learning and process modeling.
\end{IEEEbiography}

\begin{IEEEbiography}[{\includegraphics[width=1in,height=1in,clip,keepaspectratio]{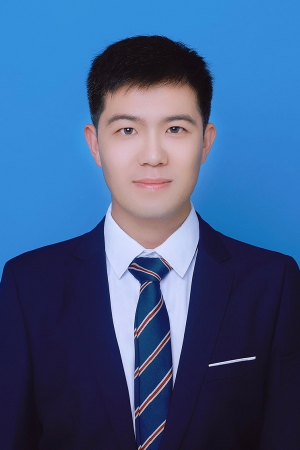}}]{Yanlin He} received the B.S. and Ph.D. degree in College of Information Science and Technology from Beijing University of Chemical Technology, in 2011 and 2016, respectively. 

Now he is a professor in College of Information Science and Technology from Beijing University of Chemical Technology, China. His research interests include fault diagnosis, soft sensor, computational intelligence, and machine learning.
\end{IEEEbiography}

\begin{IEEEbiography}[{\includegraphics[width=1in,height=1in,clip,keepaspectratio]{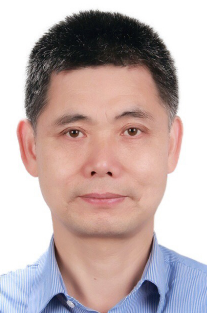}}]{Qunxiong Zhu} received the Ph.D. degree in College of Information Science and Technology from Beijing University of Chemical Technology, in 1996. 

Now he is a professor in College of Information Science and Technology from Beijing University of Chemical Technology, Beijing, China. His research interests include computational intelligence, artificial intelligence, machine learning, fault diagnosis and process modeling.
\end{IEEEbiography}

\end{document}